# The Lorentz Condition is Equivalent to Maxwell Equations

Edmund A. Di Marzio
Bio-Poly-Phase, 14205 Parkvale Road, Rockville, MD 20853
Received 26 November 2008

It is shown that the Lorentz condition which is a conservation law on the electromagnetic four-vector-density $A^\mu$, plus the Lorentz transformation, taken together, are equivalent to the microscopic Maxwell's equations.

Jackson in his book "Classical Electrodynamics" derives Eqs 1 and 2 from the microscopic Maxwell's equations[1]. They are, in Gaussian units,

$$\frac{\partial A^\mu}{\partial x^\mu} = 0 \tag{1}$$

$$\Box A^\mu = \frac{4\pi}{c} j^\mu \tag{2}$$

Also, it is straightforward to derive the microscopic Maxwell's equations from Eqs. 1 and 2. This means that the above equations are equivalent to the microscopic Maxwell's equations. Eq. 1 is the Lorentz condition (LC) and $\Box$ is (minus) the invariant d'Alembertian.

$$\Box = \left( \frac{\partial^2}{\partial (x^0)^2} - \frac{\partial^2}{\partial (x^1)^2} - \frac{\partial^2}{\partial (x^2)^2} - \frac{\partial^2}{\partial (x^3)^2} \right) \tag{3}$$

where, $(x^0, x^1, x^2, x^3) = (ct, x, y, z)$. The electromagnetic four-vector-density is $A^\mu$, $(A^0, A^1, A^2, A^3) = (\phi, A^1, A^2, A^3)$. The charge-current-density is $j^\mu$, $(j^0, j^1, j^2, j^3) = (c\rho, j^1, j^2, j^3)$, where ρ is charge density. One should also note that the LC is a conservation law on the four-vector density $A^\mu$. Eq. 1 is also invariant to the Lorentz gauge.

A more precise statement of the claim of this paper is that the LC plus the Lorentz transformation (LT), taken together, have the same formal structure (syntax) as the microscopic Maxwell's equations.

Our derivation uses the fact that the d'Alembertian, is an invariant operator so that when it operates on a any vector the result is always a vector.

We now argue that (the "one") Eq. 1 implies (the four) Eqs. 2. Since $A^\mu$ is a function of the time and space variables we can operate on it by the invariant d'Alembertian to obtain

$$\Box A^\mu = \frac{4\pi}{c} k^\mu \tag{4}$$

where $k^\mu$ is some four-vector-density.

We make the following 3 points. First, $k^\mu$ and $j^\mu$ have the same dimensions. Second, the divergence $\dfrac{\partial}{\partial x^\mu}$ commutes with $\Box$ so we have from Eqs 2 and 4

$$\frac{4\pi}{c}\frac{\partial j^\mu}{\partial x^\mu} = \frac{\partial}{\partial x^\mu}\Box A^\mu = \Box \frac{\partial A^\mu}{\partial x^\mu} = 0 \tag{5}$$

$$\frac{4\pi}{c}\frac{\partial k^\mu}{\partial x^\mu} = \frac{\partial}{\partial x^\mu}\Box A^\mu = \Box \frac{\partial A^\mu}{\partial x^\mu} = 0 \tag{6}$$

which are conservation laws on the four-vector-densities $j^\mu$ and $k^\mu$. Third, we also have

$$j^0 dxdydz = (j^0)' dx'dy'dz' \tag{7}$$

$$k^0 dxdydz = (k^0)' dx'dy'dz' \tag{8}$$

where the primed coordinates are connected to the unprimed coordinates by a LT. For charge this is the statement that the charge is an invariant so that, no matter with what speed we approach a charge distribution, the amount of charge does not change. It is a matter of experience that charge is both a conserved quantity and an invariant quantity.

Since, as we see, there is nothing that distinguishes $k^\mu$ from $j^\mu$ they are identical.

At this point a discussion of the difference between syntax and semantics is needed. Consider the heat diffusion and the particle diffusion equations. The syntax is the same; that is to say the equations have the same form. But the semantics, or meaning is much different. What we have done above is show that we can develop from the LC alone a set of equations that are identical in form to the microscopic Maxwell's equations. That is to say, we have obtained equations with the same syntax as Maxwell's equations. Our derivation also shows that any 4-vector whatsoever for which the 4-divergence is zero has the same syntax as Maxwell's equations.

Note that we have not derived Maxwell's equations from first principles. A complete derivation of Maxwell's equations from first principles requires that prior meaning be assigned to $A^\mu$. Then Maxwell's equations would be derivative both by syntax and by meaning. But how can we cloth them with meaning? The experiments of Faraday and Ampere and the genius of Maxwell have long ago clothed the Maxwell equations with meaning. At this point in time we can say that we have derived the Maxwell equations from the LC only by appropriating this meaning.

We make the following observations.

[1] Various relativists have observed that the constraint of covariance is a severe constraint on the possible forms of the laws of electricity and magnetism. We give 3 examples.

Misner, Thorne and Wheeler[2] show that the homogeneous equation div**B**=0 plus relativity implies the homogeneous equation curl**E** +∂**B**/c∂t =0.

Using relativity Kobe[3] is able to derive Maxwell's equations using only Coulomb's Law and conservation of electric charge.

Anderson[4] observes that the assumption that the electromagnetic field be an asymmetric tensor and also that the equations be linear, implies that in free space the equations $F^{\mu\nu}{}_{,\mu} = 0$, $F_{[\mu\nu,\rho]} = 0$ (his notation) are the only possible ones.

Perhaps then our derivation is not so surprising.

[2] We note the Aharonov-Bohm papers[5] which show a deep connection of $A^\mu$ to quantum mechanics. The quantum mechanical wave function is dependent on $A^\mu$ even when $A^\mu$ is a constant so that **E** and **B** are equal to zero; **B** = curl**A**, **E**= -gradφ-∂**A**/c∂t. At a bare minimum this means that $A^\mu$ has a **physical reality** above and beyond the formulas relating it to **E** and **B** since it can exist even when **E** and **B** do not. This requires us to seek the meaning of the electromagnetic 4-vector not from **E** and **B** but from something more primal. Since both the LT and the LC were each required in our derivation of Maxwell's equations we need to understand each of them at a deeper level.

[3] Some time ago an attempt was made to derive the LT and the LC from more fundamental considerations and thereby obtain Maxwell's equations in both their syntax and their semantics.

First, the LT was derived in a manner that took no note of Maxwell's equations, nor of the fact that the speed of light is a constant, nor even that light existed[6]. The arbitrary constant c with the dimensions of velocity that appears in the derivation is given the numerical value of the speed of light because, as is easily shown, the LT implies that something traveling with the speed c in one coordinate system travels with the speed c in every coordinate system. Light has this exact property. This means that although Maxwell's equations are in no way implied by the LT alone, they are subject to it, and their functional form is severely limited by the constraint of covariance. At a bare minimum it is implied that the natural structure or texture of the atoms comprising the meter sticks and clocks involves light in a most fundamental way since they were the only things assumed in our derivation. As an aside we remark that relatively few in the physics community are aware that the LT can be derived without assuming the existence of light[7].

Second, the fact that the LC has the form of a conservation law forces us to search for a set of basic equations with a symmetry whose associated conservation law is the LC. It is an easy matter to show that assuming two unique speeds is not a covariant concept since on performing a boost only the speed c will remain c, the other speed will change; similarly for n distinct unique speeds. Occam's razor forces us to the bold, but simple view that there exists only one speed in these basic equations, that of light in a vacuum. It was hypothesized[8] that there exists a fundamental field $\rho^\mu(x^\nu, \mathbf{\Omega})$ from which all the various laws of physics are derivable. At each space-time point this field which is a null 4-vector-density travels in each of the directions **Ω,** always with the speed of light**.** The conservation law on $\rho^\mu(x^\nu, \mathbf{\Omega})$ arises from the fact that the total amount of stuff in the Universe is a constant. This means that $\mathbf{Q}^\mu = \int \rho^\mu(x^\nu, \mathbf{\Omega}) \, d\mathbf{\Omega}$ obeys a conservation law and by identifying it with the electromagnetic 4-vector $A^\mu$, Eq. 1 obtains[8].

Even before obtaining the basic equations governing the behavior of $\rho^\mu(x^\nu, \mathbf{\Omega})$ we see 3 concordances with reality. First, the concept is manifestly covariant; null vectors transform to null vectors. Second, the concept of mass arises from the simple fact that the sum of null vectors is not a null vector. Third the $\rho^\mu(x^\nu, \Omega)$ field serves to transfer length from point to point so that Weyl geometry reduces to Riemannian geometry.

In order for the $\rho^\mu(x^\nu, \Omega)$ field (particles, electric field, gravity) to persist at a given place stuff must scatter stuff, otherwise it would explosively leave the place with the speed of light. Because action at a distance is meaningless in relativity, the principle of contiguous action requires that $\rho^\mu(x^\nu, \Omega_i)$ scatters $\rho^\mu(x^\nu, \Omega_j)$ at the same space-time point. We argued for the uniqueness of a simple dot product. Equations for the space-time variation of $\rho^\mu(x^\nu, \Omega)$ were obtained[8]. This procedure will be meaningful only if various measures of $\rho^\mu(x^\nu, \Omega)$ can be identified with the conserved quantities of physics such as mass, energy, and spin. This has been done, for a toy world of (1+1) dimensions[6,9,10] but not yet for our real world of (3+1) dimensions.

[4] The question can be asked whether Eqs.1 plus 2 are an overdetermined system since we have more equations than variables. But because Eq. 2 has been derived from Eq 1 it is obvious that we have a consistent set.

How can one equation (the Lorentz condition) lead to many equations (Maxwells Equations)? We stress here that the Lorentz condition is in fact many equations. This is because it embodies both the Lorentz transformation and the condition formulated by Einstein that, in special relativity, the laws of physics have the same form in all coordinate systems moving with constant relative velocity . The transformation from the unprimed $A^\mu$ and (*ct, x. y, z*) to the primed $A^\mu$ and (*ct, x. y, z*) involves six continuous parameters, the three components of the velocity and the three angles of rotation of the coordinate system. So there is a sense for which the Lorentz condition really corresponds to an infinite number of equations.

[5] Another question has to do with causality. Eq. 2 can be solved by Green function techniques and the result is that there are both advanced and retarded solutions. It is the retarded solution that is usually thought to correspond to reality since the response to a change in the source must occur later than the change in the source. If we know $j^\mu$ we can use Green's functions (plus boundary conditions) to determine $A^\mu$; but if we know $A^\mu$ we can use Eq.2 to determine $j^\mu$. Each implies the other. Also, there are meaningful solutions to Maxwell's equations which require both the retarded and advanced solutions. We see no reason to give $j^\mu$ priority over $A^\mu$, or $A^\mu$ priority over $j^\mu$.

[6] In limiting ourselves to the microscopic Maxwell equations (that is to say we assume **E** = **D** and **B** = **H** ) we ignore the properties of Materials. Because of causality **D** can lag **E** ; alternatively the dielectric constant can be frequency dependent and the Kramers-Kronig relations obtain. We will not try to extend our results to the full Maxwell equations here, but we expect the full Maxwell's equations to be valid in all the coordinate systems of special relativity.

[7] The Lorentz condition (Lorentz gauge) is a covariant concept. The Coulomb gauge, though useful is not a covariant concept and is not relevant to our discussion.

[8] Can our equations which are covariant to special relativity be made generally covariant? If we interpret $A^\mu$ as a vector-density rather than a vector then Eq. 1 is generally covariant[11]. Equation 2 is not generally covariant. The generally covariant Maxwell equations would require a generally covariant d'Alembertian. Several candidates exist[12].

In conclusion, the wave equation, Eq. 2 has been derived from Eq. 1; it did not need to be assumed. Therefore the syntax of Maxwell's equations is derived from the Lorentz condition alone (but under the aegis of special relativity). That the Lorentz Condition alone implies the microscopic Maxwell's equations **is** a remarkable result.